**基金项目**：吉林省自然科学基金项目(YDZJ202101ZYTS149)。
项目名称:面向系统韧性提升的综合能源系统优化调度研究
起止时间:2021.01.01-2023.12.31　负责人姓名：李扬
基金项目：吉林省自然科学基金项目(YDZJ202101ZYTS149)。
Project Supported by the Natural Science Foundation of Jilin Province(YDZJ202101ZYTS149).

# 阶梯式碳交易机制下考虑需求响应的综合能源系统低碳优化调度

王利猛 [1]，刘雪梦 [1]，李扬 [1]，常铎 [2]，任星 [1]

(1.电力系统仿真控制与绿色电能新技术教育部重点实验室 (东北电力大学), 吉林省吉林市 132012；2.东北电力大学计算机学院，吉林省吉林市 132012)

**摘要**：在综合能源系统(integrated energy system，IES)的运行中，考虑到进一步降低碳排放量，提升其能源利用率，愈加优化与完善 IES 的整体运行，提出了一种在阶梯式碳交易机制下考虑需求响应的综合能源系统优化调度策略。首先从需求响应(demand response，DR)角度出发，考虑到多种能源之间具备协同互补与灵活转换的能力，由此引入电-气-热的横向时移与纵向互补替代策略并构建 DR 模型；其次从生命周期评估的角度出发，阐述碳排放权初始配额模型，并对其加以修正，然后引入阶梯式碳交易机制，这种机制对 IES 的碳排放有一定程度上的约束作用；最后以能源购买成本、碳排放交易成本、设备维护成本、需求响应成本之和最小化为目标，且在考虑安全约束的情况下构建低碳优化调度模型。此模型将原问题利用 Matlab 软件转化为混合整数线性问题，并使用 CPLEX 求解器对模型进行优化求解。算例结果表明，在阶梯式碳交易机制下考虑碳交易成本和需求响应，使得 IES 的运行总成本下降了 5.69%，碳排放量降低了 17.06%，此模型显著提高了 IES 的可靠性、经济性和低碳性。
**关键词**：阶梯式碳交易；综合能源系统；需求响应；横向时移与纵向互补替代；低碳优化

**Low-carbon optimal dispatch of integrated energy system considering demand response under the tiered carbon trading mechanism**

WANG Limeng[1], LIU Xuemeng[1], LI Yang[1], CHANG Duo[2], REN Xing[1]

(1. Key Laboratory of Modern Power System Simulation and Control & Renewable Energy Technology (Northeast Electric Power University), Ministry of Education, Jilin 132012, Jilin Province, China; 2. School of computer science, Northeast Electric Power University, Jilin 132012, Jilin Province, China)

**ABSTRACT:** In the operation of the integrated energy system (IES), considering further reducing carbon emissions, improving its energy utilization rate, and optimizing and improving the overall operation of IES, an optimal dispatching strategy of integrated energy system considering demand response under the stepped carbon trading mechanism is proposed. Firstly, from the perspective of demand response (DR), considering the synergistic complementarity and flexible conversion ability of multiple energy sources, the lateral time-shifting and vertical complementary alternative strategies of electricity-gas-heat are introduced and the DR model is constructed. Secondly, from the perspective of life cycle assessment, the initial quota model of carbon emission allowances is elaborated and revised. Then introduce a tiered carbon trading mechanism, which has a certain degree of constraint on the carbon emissions of IES. Finally, the sum of energy purchase cost, carbon emission transaction cost, equipment maintenance cost and demand response cost is minimized, and a low-carbon optimal scheduling model is constructed under the consideration of safety constraints. This model transforms the original problem into a mixed integer linear problem using Matlab software, and optimizes the model using the CPLEX solver. The example results show that considering the carbon trading cost and demand response under the tiered carbon trading mechanism, the total operating cost of IES is reduced by 5.69% and the carbon emission is reduced by 17.06%, which significantly improves the reliability, economy and low carbon performance of IES.

中图分类号：

## 0 引言

进入 21 世纪以来，以数字化、网络化、智能化、绿色化为核心特征的新一轮工业变革，正在突破资源和环境的瓶颈，极大地提高了可持续发展能力，对能源系统提出了更高的要求。与此同时，在能源互联网建设和全球环境保护的大趋势下，我国提出了建立碳交易市场等政策机制，中国的能源革命加速了实现全球碳中和的承诺[1]。然而，传统的能源系统是独立规划和运行的，无法利用各种能源形式的优势互补，这极大地限制了能源的综合利用效率[2-3]。于是近年来兴起的综合能源系统(integrated energy system，IES)将信息与物理融合、多能联供和多种热、电、气储能技术相结合，实现多能转换、储存和消纳[4-6]，因此在当前数字化和能源革命的背景下，IES 已成为必然选择，是节能降碳的必要路径。

综合能源系统作为一种技术在增加清洁能源的份额和减少二氧化碳排放方面发挥着重要作用[7-8]。文献[9-10]阐释了碳交易机制，并将其与 IES 相结合，进而分析了碳交易成本在系统运行方面的重要作用；文献[11]根据核电厂、火电厂和风电厂的实际无偿碳排放权初始配额，并以火电厂的实际碳排放量为基础计算碳交易的成本，这将实现经济效益与低碳排放之间的有效平衡；文献[12]推荐了一种基于基准方法的虚拟发电厂碳交易机制，该机制根据可再生能源单位发电量分配碳源。在文献[13]中，碳排放的处理构建了一个以最小碳排放为目标的多目标模型。虽然这些方法在减少碳排放方面有一定的效果，但不能同时促使系统成本下降。为此，需要引入更加合理愈加完善的碳交易机制。

与此同时，为提高用户对可再生能源的使用，提高系统的经济效益，应对需求侧加以考虑。通过碳排放权交易、需求侧管理等机制对能源行业节能减排进行管理，被视为促进可持续发展的重要举措。文献[14]为提高能源利用率，基于热负荷的模糊性和滞后性，建立了基于价格弹性矩阵的电、气、热负荷的需求响应（demand response，DR）模型；文献[15]将电力负荷分为可减少、可重新分配和可替换三种类型，并根据响应量统一计划平衡成本，并采用饱和和分散指标来衡量用户满意度。文献[16]通过扩展传统的电力时移 DR，提出了考虑电、气、热三种时移需求响应形式的综合能量概率优化调度模型。文献[17-18]从环境经济调度和能量转移的角度，在多能源系统的最优运行下，考虑了基于价格的综合需求响应（integrated demand response，IDR），文献[19]进一步将电、气、热负荷的时移 DR 和可中断 DR 结合起来，并对 IDR 方法在能量管理中的作用进行了扩展和分析，但在研究中未提及用户侧多种能量需求的灵活转换和交互响应。

以上分析表明，大多数优化配置研究只考虑碳排放权交易或需求侧管理，在同时考虑碳排放权交易和需求侧管理方面相对薄弱。然而在优化电网时，同时考虑排放交易

和需求方管理，可以在低碳排放和经济效率之间进行最佳权衡。随着碳排放权交易的引入，传统的以经济效率为基础的规划模型正在向充分考虑低碳足迹的系统转变。引入以碳排放权交易为基础的需求侧管理，既能提高用户对可再生能源的利用，又能降低储能容量，提高系统的经济效益。

针对目前的形势和挑战，本文将进行如下研究。首先阐述碳排放权初始配额模型，并对其加以修正，然后引入阶梯式碳交易机制，限制 IES 的碳排放；其次从需求响应角度出发，引入电-气-热的横向互补替代与纵向时移策略；最后构建以 IES 运行总成本最小化为目标，并在考虑安全约束的情况下构建低碳优化调度模型；最后通过 5 种不同情景分析验证模型的正确性和有效性。

## 1 构建含电-气-热需求响应的 IES

IES 是一种复杂的结构体系，内含多种的能源形式。IES 可以利用电、热、气之间的互补协同效应，提高能源效率，满足客户对不同类型能源的交错需求，同时确保持续可靠的能源供应[20-21]。本文在阶梯式碳交易机制下，构建了含有电、气、热需求响应的 IES 的框架体系如图 1 所示。

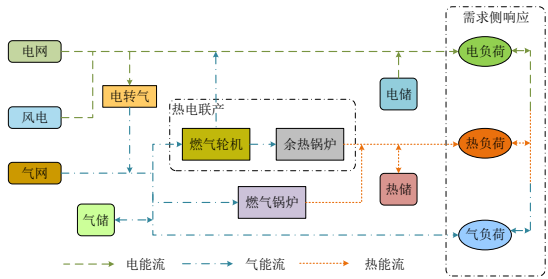

**图 1　IES 的框架体系**

**Fig. 1　integrated energy system framework system**

IES 的能源供应侧包含上级电网、上级气网和风电设备；能源转换设备包含电转气(P2G)、燃气锅炉(GB)、燃气轮机(GT)、余热锅炉(WHB)。热电联产由燃气轮机(GT)和余热锅炉(WHB)组成，采用热电联产耦合模式运行，可适应不同的系统条件；能源存储是由电储、气储和热储构成；电负荷、气负荷以及热负荷三部分共同构成系统的能源需求侧响应。需求响应的引入可以平滑负荷曲线的波动，实现功率和热量的交叉耦合，减少峰谷差，降低运营成本。

## 2 基于横向时移与纵向互补替代策略的需求响应模型

在能源互联网快速发展的背景下，电、气、热等多种能源之间的协同互补与灵活转换，可以为需求方参与系统优化提供更多机会。本文从需求响应角度出发，引入电-气-热的横向时移与纵向互补替代的策略，即运营商可以根据能源价格的变化制定不同的调度计划，进一步满足多样化能源需求，以确保 IES 的经济、灵活和高效的运作。参照传统电力需求分类，根据交互响应特性，将 IES 中的需求响应负荷（电、气、热）扩展为三种负荷类型，即固定型负荷，可转移型负荷，可代替型负荷。

$$P_{k,load}(t) = P_{k,load}^{s}(t) + P_{k,load}^{p}(t) + P_{k,load}^{c}(t) \quad (1)$$

式中 $k$ 表示负荷类型；$P_{k,load}(t)$ 表示在 $t$ 时段下第 $k$ 种负荷的功率，单位：kW；$P_{k,load}^{s}(t)$、$P_{k,load}^{p}(t)$、$P_{k,load}^{c}(t)$ 分别表示在 $t$ 时段下第 $k$ 种负荷的固定型负荷、可转移型负荷、可代替型负荷的功率，单位：kW。

### 2.1 固定型负荷

固定型负荷是指在正常运行中需要完全满足的部分负荷，如照明、官方负荷和大多数商业负荷。固定负荷基本不参与需求响应，但在发生事故或突发事件时也可以部分切断，以确保系统的安全。当发生这种情况时，运营商需要向用户提供补偿。

### 2.2 可转移型负荷

可转移型负荷是指在能源总需求大致不变的情况下，根据能源价格的变化，能够及时进行横向移位和调整，实现负荷调峰的部分负荷。可转移型负荷主要由居民用户和部分工业用户组成，涉及洗衣机、热水器等设备和部分工业设备。

### 2.3 可替代型负荷

可替代型负荷是指可以用其他能源形式替代以满足能源需求的部分负荷；这主要影响到一些住宅和商业用户，并涉及空调、炉灶和供暖设备等设备。由于更换 DR 是用户自愿按照能源价格实现的，因此不会对用户日常生活的舒适度产生明显影响[22]。

可转移型负荷与可替代型负荷模型如下所示：

$$\begin{cases} P_{k,load}^{n*} = P_{k,load}^{n}(t) + \Delta P_{k,load}^{n}(t) \\ \Delta P_{k,load}^{n}(t) = v_{k,in}^{n} P_{k,in}^{n}(t) - v_{k,out}^{n} P_{k,out}^{n}(t) \\ v_{k,in}^{n} + v_{k,out}^{n} = 1 \\ \sum_{t=1}^{T} \Delta P_{k,load}^{n}(t) = 0 \\ P_{k}^{n,\min} \leq \Delta P_{k,load}^{n}(t) \leq P_{k}^{n,\max} \end{cases} \quad (2)$$

式中：$n$ 表示负荷的不同类型。当 $n$=p 时，$P_{k,load}^{p*}$ 表示第 $k$ 种负荷 $t$ 时段的可转移型负荷参与需求响应之后的功率，$\Delta P_{k,load}^{p}(t)$ 表示第 $k$ 种负荷 $t$ 时段的可转移型负荷参与需求响应的功率；当 $n$=c 时，$P_{k,load}^{c*}$ 表示第 $k$ 种负荷 $t$ 时段的可替代型负荷参与需求响应之后的功率，$\Delta P_{k,load}^{c}(t)$ 表示第 $k$ 种负荷 $t$ 时段的可替代型负荷参与需求响应的功率；$v_{k,in}^{n}$、$v_{k,out}^{n}$ 分别表示在 $t$ 时段下第 $k$ 种负荷的转入、转出参数，两者皆为二进制参数；$P_{k,in}^{n}(t)$、$P_{k,out}^{n}(t)$ 分别表示第 $k$ 种负荷在 $t$ 时段转入、转出功率；$P_{k}^{n,\min}$、$P_{k}^{n,\max}$ 分别表示第 $k$ 种负荷参与到需求响应的下限值、上限值。功率的单位：kW。

### 2.4 需求响应模型

综上所述，DR 模型可以通过多种需求响应方式来改变能源需求，具体如下：

$$\begin{aligned} P_{k,load}^{*} &= P_{k,load}(t) + \Delta P_{k,load}(t) \\ &= P_{k,load}(t) + \Delta P_{k,load}^{p}(t) + \Delta P_{k,load}^{c}(t) \end{aligned}$$
(3)

式中：$P_{k,load}^{*}$ 表示在 $t$ 时段下第 $k$ 种负荷参与需求响应之后的功率，单位：kW；$\Delta P_{k,load}(t)$ 表示在 $t$ 时段下第 $k$ 种负荷参与需求响应的功率，单位：kW。

## 3 基于生命周期评估的阶梯型碳交易机制

综合能源系统可以提高能源效率，在 IES 中合理配置低碳设备可以有效减少碳排放，实现碳达峰和碳中和。碳排放主要来自能源开采、能源运输和使用等各个阶段，然而现有的综合能源系统碳排放研究大多集中在运行阶段或其中的单个模块上，缺乏对综合能源系统所有阶段碳排放的分析和预测。因此本文从生命周期评估的角度出发，所考虑的能源包括煤电、风电机组、天然气。将能源和储能设备在生产、运输、运行和废弃的全生命周期碳排放转化为碳排放成本纳入优化目标中，分析在考虑生命周期碳成本后，对系统优化结果的影响[23]。各个环节具体计量过程可参考文献[24]。

在国家的大力推进下，碳交易机制正逐步实施与推进，碳交易市场的发展与完善可以在很大程度上促进全国各地低碳减排的进程。政府或相关机构会为 IES 的每个碳排放源发放无偿的碳排放权配额。如果生产厂商的实际二氧化碳排放量小于政府分配的配额，它可以出售多余的碳排放权配额来产生收入；相反，生产厂商必须购买碳排放权配额来抵消多余的二氧化碳排放。碳交易机制其模型主要由 3 个部分组成：碳排放权初始配额模型、实际碳排放模型以及阶式型碳交易成本计算模型。

### 3.1 碳排放权初始配额模型

IES 中的碳排放主要来源于上级购电、GT、GB 和需求侧气负荷。碳排放权初始配额模型如下：

$$\begin{cases} E_{IES} = E_{e,buy} + E_{GT} + E_{GB} + E_{g,load} \\ E_{e,buy} = \sigma_{e} \sum_{t=1}^{T} P_{e,buy}(t) \\ E_{GT} = \sigma_{h} \sum_{t=1}^{T} (\sigma_{e,h} P_{GT,e}(t) + P_{GT,h}(t)) \\ E_{GB} = \sigma_{h} \sum_{t=1}^{T} P_{GB,h}(t) \\ E_{g,load} = \sigma_{g,load} \sum_{t=1}^{T} P_{g,load}(t) \end{cases} \quad (4)$$

式中：$E_{IES}$、$E_{e,buy}$、$E_{GT}$、$E_{GB}$、$E_{g,load}$ 分别为综合能源系统、系统向上级购电、燃气轮机、燃气锅炉、需求侧气负荷的无偿碳排放权配额；$\sigma_{e}$、$\sigma_{h}$ 分别为产生单位电量、单位热量所获得的无偿碳排放权配额；$\sigma_{g,load}$ 为消耗单位气负荷的无偿碳排放权配额。无偿碳排放权配额单位：$kgCO_2$。$\sigma_{e,h}$ 为燃气轮机发电量向发热量的折算参数；$P_{e,buy}(t)$ 表示在 $t$ 时段下系统向上级购电的功率；$P_{GT,e}(t)$、$P_{GT,h}(t)$ 分别表示在 $t$ 时段下燃气

轮机电功率、热功率的供给量；$P_{GB,h}(t)$表示在 $t$ 时段下燃气锅炉热功率的供给量；$P_{g,load}(t)$表示在 $t$ 时段下气负荷的消耗量；$T$ 为系统的一个调度周期。功率的单位：kW。

### 3.2 实际碳排放模型

系统在电转气的过程中会消耗部分$CO_2$，同时需求侧的气负荷在运行时也会有大量$CO_2$产生，如若将其考虑进来，本文则需对原有模型进行修正。由此得出，实际碳排放模型如下：

$$\begin{cases} E_{IES,a} = E_{e,buy,a} + E_{GBGT,a} + E_{g,load,a} - E_{P2G,a} \\ E_{e,buy,a} = \sum_{t=1}^{T}\left(a + bP_{e,buy}(t) + cP_{e,buy}^2(t)\right) \\ E_{GTGB,a} = \sum_{t=1}^{T}\left(d + eP_{GTGB}(t) + fP_{GTGB}^2(t)\right) \quad (5) \\ P_{GTGB}(t) = P_{GT,e}(t) + P_{GT,h}(t) + P_{GB,h}(t) \\ E_{g,load,a} = \delta \sum_{t=1}^{T} P_{g,load}(t) \\ E_{P2G,a} = \theta \sum_{t=1}^{T} P_{P2G,g}(t) \end{cases}$$

式中：$E_{IES,a}$为综合能源系统的实际碳排量；$E_{e,buy,a}$为系统向上级购电的实际碳排量；$E_{GTGB,a}$为燃气锅炉与燃气轮机两部分的实际碳排量，$P_{GTGB}(t)$表示其在 $t$ 时段下的等效输出功率；$E_{g,load,a}$为供应侧气负荷的实际碳排量，$\delta$为单位气负荷的等效碳排放参数，具体值可参考文献[25]；$E_{P2G,a}$为P2G设备在电转气的过程中吸收的$CO_2$量，$P_{P2G,g}(t)$表示其在 $t$ 时段下输出的天然气功率，$\theta$为此过程中吸收$CO_2$的参数；$a$、$b$、$c$ 为煤电机组的碳排放参数；$d$、$e$、$f$ 为消耗天然气的机组其碳排放参数。碳排放量和功率的单位分别为 kg 和 kW。

### 3.3 阶梯式碳交易成本计算模型

由IES总碳排量以及无偿碳排放权配额即可算出参与到碳交易市场的碳排放交易份额：

$$E_{IES,c} = E_{IES,a} - E_{IES} \quad (6)$$

式中：$E_{IES,c}$为系统参与到碳交易市场的碳排放交易份额；$E_{IES}$为系统获得的无偿碳排放权配额；$E_{IES,a}$为系统的实际碳排量。单位均为：kg。

阶梯式碳交易成本计算模型在分层排放权交易下，将二氧化碳排放量划分为不同的范围。二氧化碳排放量越高，交易成本越高，交易价格也就越高[26-27]。构建阶梯式碳交易成本分段线性计算模型，指定若干排放区间。当$E_{IES,c}$为负时，意味着系统排放的碳量低于标准量，则可以在相应的交易中心以规定的价格出售，超出的份额可以获得一定的补贴，碳排放区间越小对应的碳交易价格越高；反之亦然。求解时，碳交易成本可表示为如下的分段函数：

$$C_{CO_2} = \begin{cases} \lambda E_{IES,c}, & E_{IES,c} \leq d \\ \lambda(1+\alpha)(E_{IES,c} - d) + \lambda d, & d \leq E_{IES,c} \leq 2d \\ \lambda(1+2\alpha)(E_{IES,c} - 2d) + \lambda(2+\alpha), & 2d \leq E_{IES,c} \leq 3d \\ \lambda(1+3\alpha)(E_{IES,c} - 3d) + \lambda(3+3\alpha), & 3d \leq E_{IES,c} \leq 4d \\ \lambda(1+4\alpha)(E_{IES,c} - 4d) + \lambda(4+6\alpha)d & 4d \leq E_{IES,c} \leq 5d \\ \lambda(1+5\alpha)(E_{IES,c} - 5d) + \lambda(5+10\alpha)d & 5d \leq E_{IES,c} \end{cases}$$

(7)

式中：$C_{CO_2}$表示阶梯式碳交易成本，单位：元；$\lambda$表示碳交易的基准价格，单位：元/kg；$d$ 表示碳排放量区间长度，单位：kg；$\alpha$表示价格增长幅度。

## 4 阶梯式碳交易机制下考虑需求响应的综合能源系统低碳优化调度模型

### 4.1 目标函数

本文以IES的运行总成本$C$为目标函数，包括能源购买成本$C_{buy}$、碳交易成本$C_{CO_2}$、需求响应补偿成本$C_{dr}$、设备维护成本$C_{em}$，具体表示如下：

$$C_{min} = (C_{buy} + C_{CO_2} + C_{em} + C_{dr}) \quad (8)$$

（1）能源购买成本$C_{buy}$。
$$C_{buy} = \sum_{t=1}^{T} \alpha_t P_{e,buy}(t) + \sum_{t=1}^{T} \beta_t P_{g,buy}(t) \quad (9)$$

式中：$P_{e,buy}(t)$、$P_{g,buy}(t)$分别表示在$t$时段下的购电量以及购气量，单位分别为 kW/h 和 m³；$\alpha_t$、$\beta_t$分别表示在$t$时段下的电价以及气价，单位为元。

（2）阶梯式碳交易成本$C_{CO_2}$见式(7)。

（3）需求响应补偿成本$C_{dr}$。
$$C_{dr} = \sum_{k=1}^{3}\sum_{t=1}^{T}\left(\mu_p |\Delta P_{k,load}^p(t)| + \mu_c |\Delta P_{k,load}^c(t)|\right) \quad (10)$$

式中：$\mu_p$、$\mu_c$分别表示可转移型负荷、可替代型负荷参与需求响应的单位补偿系数。

（4）设备维护成本$C_{em}$。
$$C_{em} = \sum_{t=1}^{T}\sum_{n=1}^{N}\varpi_n P_n^t \quad (11)$$
式中：$N$为维修设备总数；$P_n^t$和$\varpi_n$是在t时段内的输出功率和维护价格，单位分别为kW和元。

### 4.2 约束条件

#### 4.2.1 风电机组出力约束

$$0 \leq P_{DG}(t) \leq P_{DG}^{max} \quad (12)$$

式中：$P_{DG}(t)$表示在t时段下的风电输出功率；$P_{DG}^{max}$为风电输出功率最大值。单位均为kW。

#### 4.2.2 储能运行约束

采用储能系统的通用模型对电、热储两类储能设备进行建模，此类储能模型已具备大量文献研究，文本在此不再赘述，具体模型见文献[25]。

#### 4.2.3 用户满意度约束

考虑用户满意度约束，以避免对用户的工作和生活造成影响。

$$I_{min} \leq I \leq 1 \quad (13)$$

$$I = \sum_{i\in\{e,g,h\}}\left(1 - \sum_{t=1}^{T}|P_{k,load}^*(t) - P_{k,load}(t)|\Big/\sum_{t=1}^{T}P_i\right)\Big/3$$

(14)

其中$I_{min}$是综合消费满意度指数(ICSI)的下限，本文将其设为0.85；$P_{k,load}(t)$，$P_{k,load}^*(t)$分别表示在$t$时段下需求响应之前和之后的负荷$k$的功率，单位kW。

#### 4.2.4 CHP运行约束

$$\begin{cases} P_{CHP,e}(t) = \varepsilon_{CHP}^e P_{g,CHP}(t) \\ P_{CHP,h}(t) = \varepsilon_{CHP}^h P_{g,CHP}(t) \\ P_{g,CHP}^{min} \leq P_{g,CHP}(t) \leq P_{g,CHP}^{max} \\ \Delta P_{g,CHP}^{min} \leq P_{g,CHP}(t+1) - P_{g,CHP}(t) \leq \Delta P_{g,CHP}^{max} \\ \omega_{CHP}^{min} \leq P_{CHP,h}(t)/P_{CHP,e}(t) \leq \omega_{CHP}^{max} \end{cases} \quad (15)$$

式中：$P_{g,CHP}(t)$表示在$t$时段下输入CHP的天然气功率；$P_{CHP,e}(t)$、$P_{CHP,h}(t)$分别表示在$t$时段下CHP输出的电功率和热功率；$\varepsilon_{CHP}^e$、$\varepsilon_{CHP}^h$分别表示CHP转换为电能和热能的效率；$P_{g,CHP}^{min}$、$P_{g,CHP}^{max}$分别为输入CHP的天然气功率最小值和最大值；$\Delta P_{g,CHP}^{min}$、$\Delta P_{g,CHP}^{max}$分别为CHP的爬坡下限和上限；$\omega_{CHP}^{min}$、$\omega_{CHP}^{max}$分别为CHP的热电比最小值和最大值。功率单位：kW。

#### 4.2.5 GB运行约束

$$\begin{cases} P_{GB,h}(t) = \phi_{GB}P_{GB,g}(t) \\ P_{GB,g}^{min} \leq P_{GB,g}(t) \leq P_{GB,g}^{max} \\ \Delta P_{GB,g}^{min} \leq P_{GB,g}(t+1) - P_{GB,g}(t) \leq \Delta P_{GB,g}^{max} \end{cases} \quad (16)$$

式中：$\phi_{GB}$表示GB的能量转换效率；$P_{g,GB}(t)$表示$t$时段下输入GB的天然气功率；$P_{g,GB}^{min}$、$P_{g,GB}^{max}$分别表示输入GB功率的最小值和最大值；$\Delta P_{g,GB}^{min}$、$\Delta P_{g,GB}^{max}$分别表示GB的爬坡下限和上限。

#### 4.2.6 功率平衡约束

（1）电功率平衡约束：

$$\begin{cases} P_{e,buy}(t) = P_{e,load}(t) + P_{e,P2G}(t) + P_{ES}^e(t) - P_{DG}(t) - P_{GT,e}(t) \\ 0 \leq P_{e,buy}(t) \leq P_{e,buy}^{max} \end{cases}$$

(17)

式中：$P_{e,load}(t)$表示在$t$时段下的电负荷；$P_{e,P2G}(t)$表示在$t$时段输入P2G的电功率；$P_{ES}^e(t)$表示在$t$时段输入电储能系统的功率；$P_{e,buy}^{max}$表示向系统上级电网购电功率的最大值。单位均为kW。

（2）气功率平衡约束：

$$\begin{cases} P_{g,buy}(t) = P_{g,load}(t) - P_{P2G,g} + P_{ES}^g(t) + P_{g,GB}(t) + P_{g,GT}(t) \\ 0 \leq P_{g,buy}(t) \leq P_{g,buy}^{max} \end{cases}$$

(18)

式中：$P_{ES}^g(t)$表示在$t$时段下输入天然气储能系统的功率；$P_{g,buy}^{max}$表示向上级天然气网购气功率的最大值。单位 kW。

（3）热功率平衡约束：

$$P_{CHP,h}(t) + P_{GB,h}(t) = P_{h,load}(t) + P_{ES}^h(t) \quad (19)$$

式中：$P_{h,load}(t)$表示在$t$时段下的热负荷，单位 kW；$P_{ES}^h(t)$表示在$t$时段下输入热储能系统的功率，单位 kW。

## 5 算例分析

为了进行案例研究，本文选取了中国东

北地区的一个综合性工业园区作为分析对象，因其对电、气、热负荷的需求是多样化的。本文系统以 24h 为一个调度周期进行仿真。IES 中的设备参数与储能参数详见附录 A 表 a1 表 a2；IES 内部风电出力与电、气、热负荷预测值详见附录 A 图 a1；分时电价、分时气价详见文献[28]；实际碳排放模型参数参考文献[29]；产生单位电量的碳排放权配额$\sigma_e$等于 0.798kg/(kW·h)、产生单位热量的碳排放权配额$\sigma_h$等于 0.385kg/(kW·h)；消耗单位气负荷的碳排放权配额$\sigma_{g,load}$等于 0.180kg/(kW·h)；碳排放量区间长度 d 等于 2000kg，价格增长幅度 $\alpha$ 等于 0.25，碳交易基价 $\lambda$ 等于 0.251 元/kg[30]；需求侧的可转移型负荷占总负荷的 10%，且可替代型负荷占总负荷的 5%。将前文所提及的低碳优化调度模型分段线性化成为线性模型，然后使用 Matlab 软件的 CPLEX 求解器对模型进行优化求解。

## 5.1 阶梯式碳交易机制下考虑需求响应的效益分析

为了验证在阶梯式碳交易机制下考虑需求响应可以事半功倍地达到预期目标，设计了以下 5 种情景进行比较和分析。

情景 1：在阶梯式碳交易机制下，不考虑需求响应，不考虑碳交易成本。

情景 2：在传统碳交易机制下，不考虑需求响应，考虑碳交易成本。

情景 3：在阶梯碳交易机制下，不考虑需求响应本，考虑碳交易成本。

情景 4：在阶梯碳交易机制下，考虑碳交易成本和需求侧负荷的横向时移。

情景 5：在阶梯碳交易机制下，考虑碳排放成本和需求侧负荷的横向时移与纵向互补替代。

每种情景系统的运行总成本和实际碳排放量见表 1。情景 1、情景 2 的电、热、气功率平衡图见图 2 图 3 所示。

**表 1 各情景优化调度结果**
**Table 1 Optimized scheduling results for each scenario**

| 情景 | 系统运行总成本/元 | 能源购买成本/元 | 碳交易成本/元 | 设备维护成本/元 | DR 补偿成本/元 | 实际碳排放量/kg |
|---|---|---|---|---|---|---|
| 1 | 30391.86 | 15810.75 | 11858.23 | 2722.88 | 0 | 23716.46 |
| 2 | 28259.20 | 16094.92 | 8989.08 | 3175.20 | 0 | 22472.70 |
| 3 | 29816.96 | 16891.83 | 9118.96 | 3806.17 | 0 | 20737.92 |
| 4 | 29298.68 | 16339.66 | 9025.78 | 3891.12 | 42.12 | 20551.57 |
| 5 | 28662.39 | 16114.30 | 8585.72 | 3888.81 | 73.56 | 19671.45 |

（1）情景 1 与情景 2 对比结果分析

情景 1 在优化调度时未考虑系统的碳交易成本，转而以能源购买成本最低为其优化调度目标，这就使得系统更倾向于购买价格较低的天然气来供能以减少其自身的能源购买成本，而大量燃烧天然气会显著增加碳排放量，只能购买碳排放权配额来抵消多余的二氧化碳排放。因此情景 1 不仅碳排放量高，系统的运行总成本也远高于其他情景。

正因为情景 2 相较于情景 1 而言考虑了系统的碳交易成本，所以才在此情景下对碳排放量有所制约。尽管天然气的购买价格较为便宜，并且低于电力的购买价格，但使用天然气会产出高量的 $CO_2$，以至于在碳交易市场上额外购买碳排放权配额所花费的成本高于系统所节省的能源购买成本。因此情景 2 的碳排放量比情景 1 低 1243.76kg，系统的运行总成本相较于情景 1 也减少了 2132.66 元。

（2）情景 2 与情景 3 对比结果分析

情景 3 相较于情景 2 而言改用阶梯式碳交易机制来计算系统的碳交易成本，而摒弃了传统的碳交易机制。首先碳交易成本作为系统运行总成本的权重，然后又由于购买碳排放权配额的价格是阶梯式递增的，所以当碳交易成本的权重增加时，激励了系统更加注重减少自身的碳排放量，所以情景 3 比情景 2 的碳排放量减少了 1734.78kg。但由于改用阶梯式碳交易机制来计算系统的碳交易成本，使得情景 3 较于情景 2 的碳交易成

本增加了 129.88 元，进而导致系统的运行总成本增加。

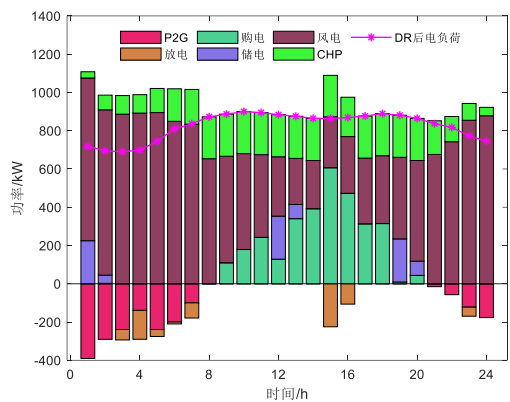

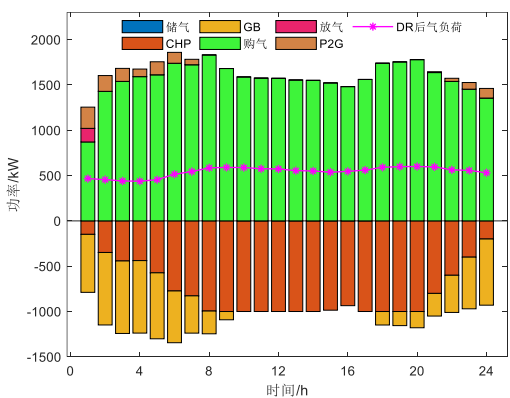

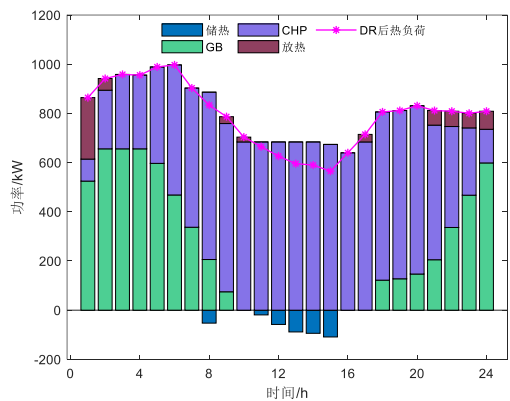

图 2 情景 1 电、气、热功率平衡图

Fig. 2 Scenario 1: Electrical, gas, and thermal power balance diagram

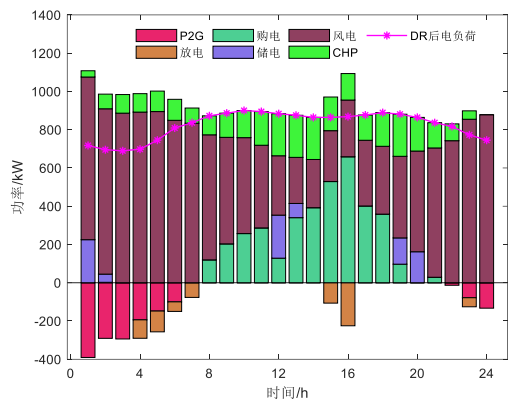

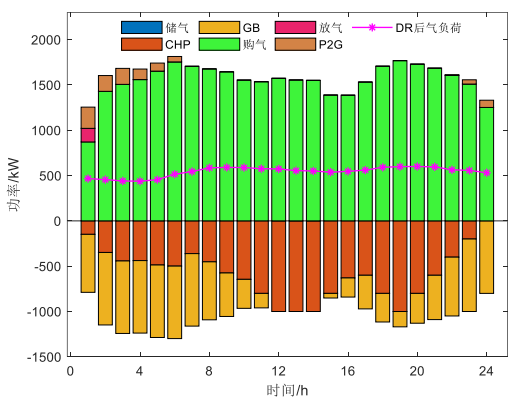

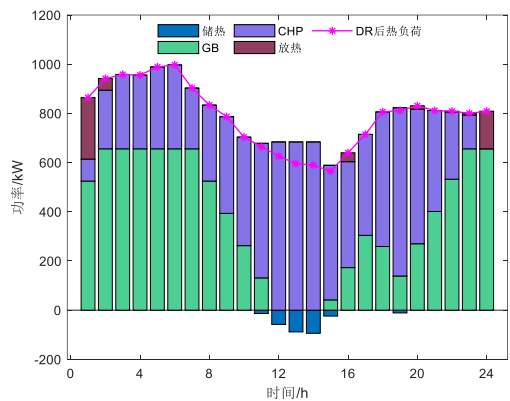

图 3 情景 2 电、气、热功率平衡图

Fig. 3 Scenario 2: Electrical, gas, and thermal power balance diagram

（3）情景 3 与情景 5 对比结果分析

情景 5 在情景 3 的基础上考虑了用户侧的需求响应，基于 DR 角度出发引入了电-气-热横向时移与纵向互补替代策略，使得用户可以根据能源价格的不同而改变他们的使用策略。例如将白天部分的负荷需求转移到夜间或其余能源价格较低的时间段，也

可以在同一时间且满足自身需求的前提下，自由选择设备以达到节能的效果。因此与情景 3 做对比，情景 5 达到了优化的效果，不但碳排放量减少了 1066.47kg，系统的运行总成本也减少了 1154.57 元，在低碳性与经济性两方面上起到了显著效果。

### 5.2 碳交易基价与区间长度影响分析

在一个考虑碳交易的系统中，因碳交易成本为目标函数的权重，所以其变化会对该系统运行总成本有着一定程度的影响。为降低碳交易成本进而减少系统运行总成本，本文做出如下研究。图 4、图 5 分别显示了碳交易基价λ、区间长度 d 与该系统的实际碳排放量及碳交易成本之间的关系曲线。

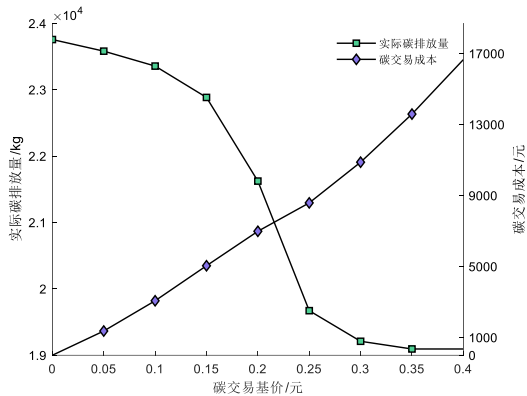

**图 4　碳交易基价对 IES 的影响**

**Fig. 4　The impact of carbon trading base prices on integrated energy systems**

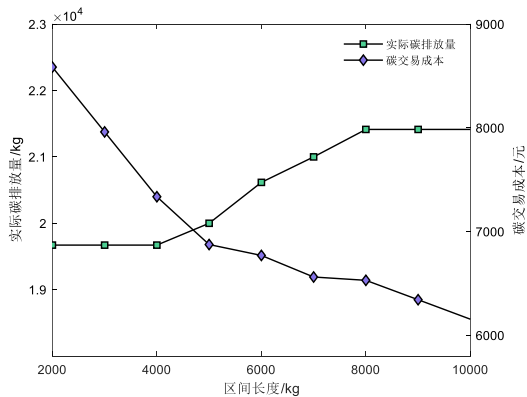

**图 5　区间长度对 IES 的影响**

**Fig. 5　The effect of interval length on integrated energy systems**

碳交易成本为系统运行总成本的权重，当λ价格上升时，碳交易成本权重随之逐渐增大，此时该系统及时地对碳价作出反应，为降低自身运行成本系统中的实际碳排放量将显著减少，但因价格的上升幅度超过实际碳排放量下降幅度，碳交易成本会逐渐增加，此时碳交易成本的增加归咎于λ价格的增加；当λ ≥ 0.35元/kg时，此时系统各部分设备的运行会保持一个平稳的状态，碳排放量达到其谷值且基本稳定，即使λ价格不断上涨，但仍无法改变系统行为，因此碳交易成本随着λ价格的上升而增加，进而导致系统的运行总成本增加。

当2000kg ≤ d < 4000kg时，因 d 的跨度空间很小，所以根据 IES 的实际碳排放量，系统需额外购买较多的碳排放权配额，但其价格处在阶梯碳价较高的区间，从而导致碳交易成本骤增，此时该系统及时地作出反应，为降低自身运行成本系统只能保持较低的碳排放量；当4000kg ≤ d < 8000kg时，因 d 的跨度空间很大，此时根据系统的实际碳排放量，其形成了在阶梯碳价较低的区间额外购买碳排放权配额的行为，使得系统的碳交易成本显著降低，但系统的实际碳排放量显著增加；当8000kg ≤ d < 10000kg时，因 d 的跨度空间过大，系统需要在碳交易市场上额外购买碳排放权配额全部处于阶梯碳价较低的区间，因为价格上升空间很小，无法改变系统自身行为，所以此时系统的实际碳排放量趋于平稳。

以上分析表明，适当的碳交易基准定价和区间长度的合理设置可以促进系统经济与低碳排放之间的协同效应。

### 5.3 基于横向时移与纵向互补替代的需求响应策略分析

在阶梯式碳交易机制下，对需求侧进行优化调度结果如下。从表 1 可以看出，与情景 3 相比，情景 4、情景 5 的碳排放量分别减少了 186.35kg、1066.47kg；就系统的运行总成本而言，情景 4、情景 5 下的系统的运行总成本较情景 3 相比分别减少 603.23 元、1237.21 元。这两种情景的需求响应结果如图 6、7 所示。三种情景的电、热功率平衡图如图 8、9、10 所示。

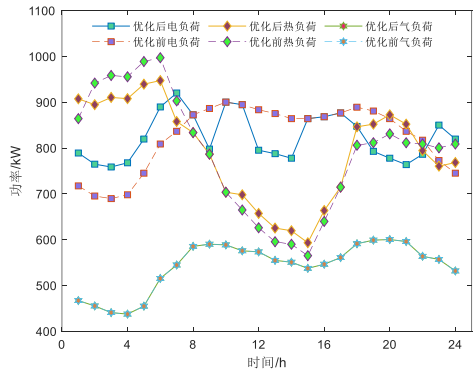

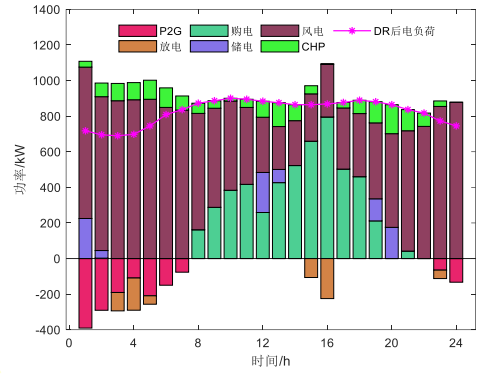

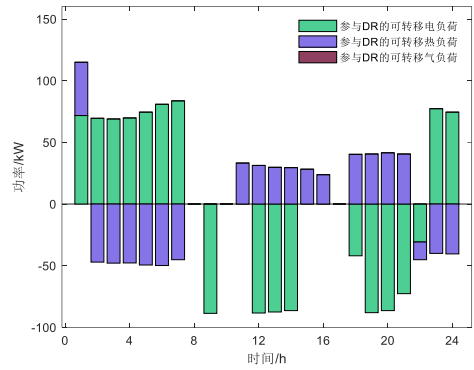

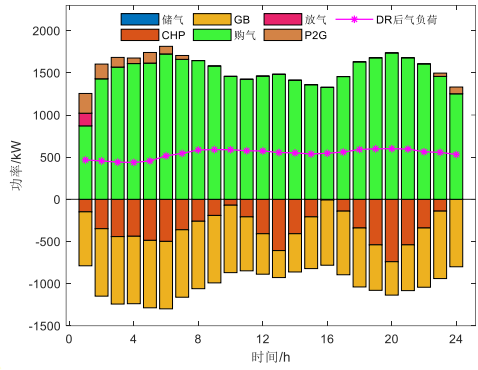

图 6 情景 4 的需求响应

Fig. 6 Demand response for scenario 4

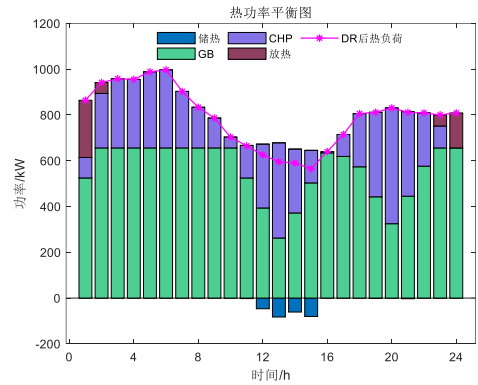

图 8 情景 3 电、气、热功率平衡图

Fig. 8 Scenario 3: Electrical, gas, and thermal power balance diagram

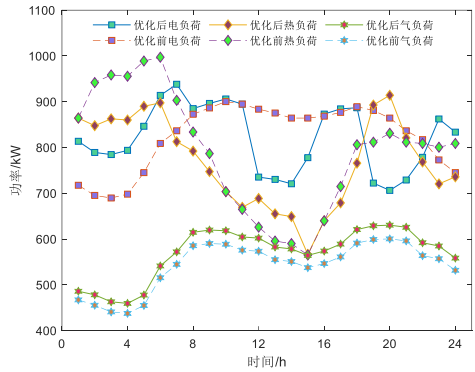

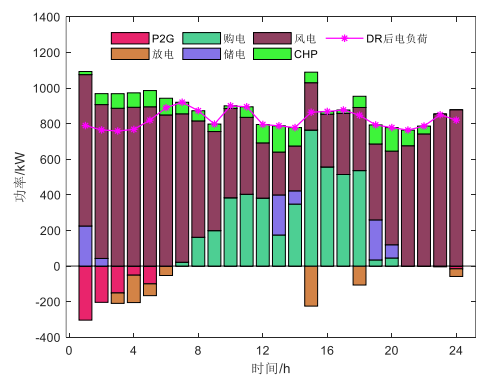

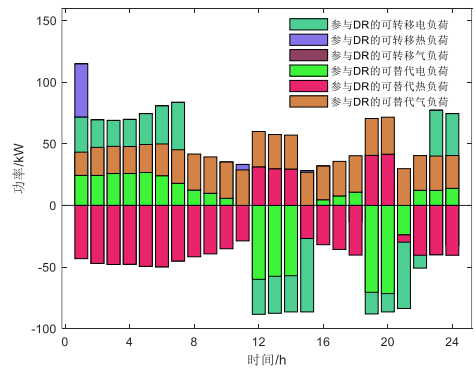

图 7 情景 5 的需求响应

Fig. 7 Demand response for scenario 5

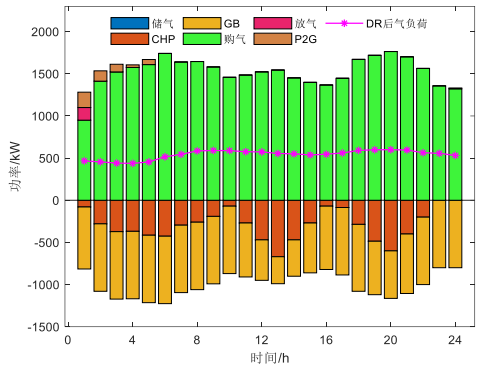

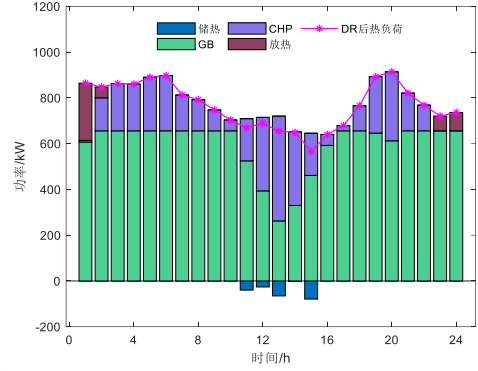

图 10 情景 5 电、气、热功率平衡图

Fig. 10 Scenario 5: Electrical, gas, and thermal power balance diagram

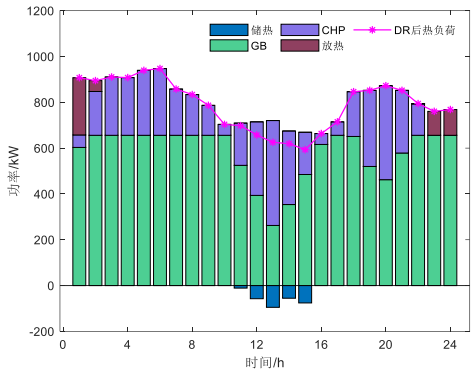

图 9 情景 4 电、气、热功率平衡图

Fig. 9 Scenario 4: Electrical, gas, and thermal power balance diagram

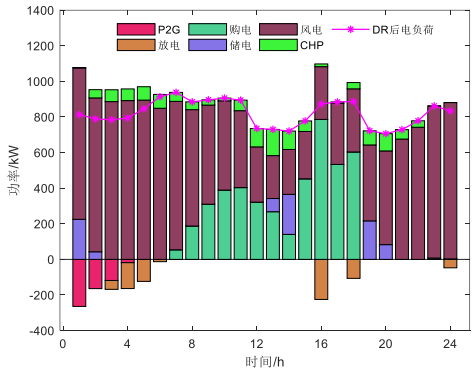

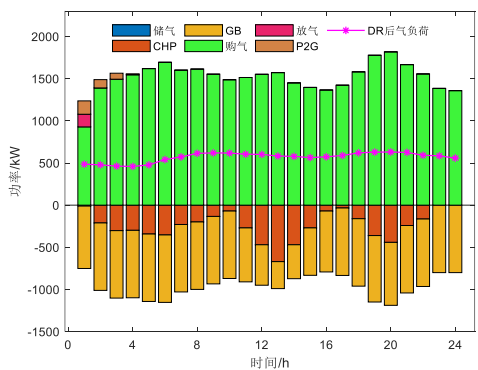

（1）情景 3 与情景 4 对比结果分析

由于夜间风电的逆峰调节特性，为减少弃风现象的发生，夜间可使用电负荷取代部分气、热负荷；与此同时，受时间不同电价不同的影响作用，IES 会将白天部分的电力负荷需求向夜间转移以提升能源利用率，进而减少系统运行总成本。

如图 2 热负荷 DR 所示，相较于白天而言用户侧对热负荷的需求量大都在夜间时段。系统的耗天然气型设备会保持着高碳排放状态以满足此时热负荷的需求，但由于此时系统对碳交易成本有所制约，因此将部分热负荷从夜间时移到白天以减少碳排放量，进而降低碳交易成本。由于考虑了需求侧负荷的横向时移，使得 CHP 出力有所减小，为更好满足用户侧对于热负荷的需求，采用 GB 进行热功率的输出，此举措可在一定程度上节省系统的运行总成本。

由于气负荷的采用横向时移策略并不会影响总体碳排量，同时采用此策略时无法避免的会有 DR 补偿反而增加成本，所以气负荷并未使用横向时移策略。

（2）情景 4 与情景 5 对比结果分析

在情景 5 的条件下，此时的需求侧采取了横向时移与纵向互补替代策略。为进一步减少弃风现象的发生，夜间可使用电负荷取代部分气、热负荷；在白天购电价格要高于购气价格，因此可使用气负荷取代部分电、热负荷。可见，DR 策略的实施能够在一定程度上改变负荷曲线，缓解能源供应压力，从而发挥出需求侧在 IES 运行中的优化潜力，

这主要源于两个方面:横向时移 DR 是将峰值时段的部分负荷功率移至较低时段,实现负荷曲线的"削峰填谷";通过纵向互补替代 DR,可以通过灵活多样的能源使用选择,实现不同能源负荷的互补替代。

# 6 结论

本文引入阶梯式碳交易机制和横向时移与纵向互补替代策略并构建了 IES 低碳优化调度模型,将 5 种典型情景下 IES 的碳排放量与系统运行总成本进行比较,并对阶梯式碳交易机制、碳交易基价λ与区间长度 d、需求侧横向时移与纵向互补替代策略对系统运行经济性和低碳性的影响进行分析,得出如下结论。

(1)本文考虑传统的碳交易机制的不足,改用阶梯式碳交易机制,其碳价的阶梯递增可以激励系统更加注重减少自身的碳排放量。碳交易基价与区间长度合理设置可以促进系统经济与低碳排放之间的协同效应。

(2)本文提出的 DR 方法可以有效实现电-气-热负荷的横向时移与纵向互补替代,在一定程度上缓解能源供应压力,有效地协调系统运行的经济性和低碳性。

(3)在阶梯式碳交易机制下考虑需求响应可以优化综合能源系统,使其达到节能减排作用效果,使得系统运行更加低碳减排且节约成本,其中在此策略下 IES 的运行总成本下降了 5.69%,碳排放量降低了 17.06%。

后续将对风电出力不确定性开展深入研究,此外也会对 IES 在多时间尺度下逐步优化,使其低碳减排、节能经济的稳步发展。

# 7 参考文献

作者简介：
王利猛(1971),男,博士,副教授,主要研究方向为电力系统监测与控制；E-mail：wlm_28@163.com；
刘雪梦(1998),女,硕士研究生,通信作者,主要研究方向为综合能源系统优化调度,E-mail：2199344907@qq.com；

## 附录 A

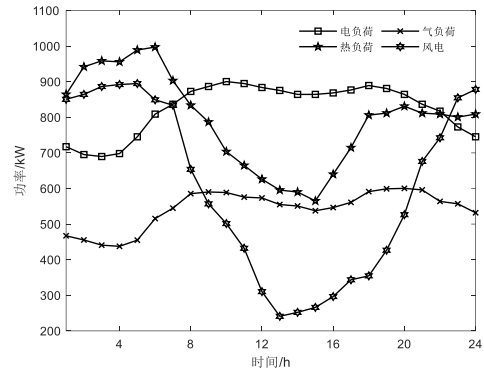

图 a1　IES 内部风电出力与电、气、热负荷预测值

Fig. a1　Integrated energy system Internal wind power output and electricity, gas and heat load forecasts

表 a1　设备参数
Table a1　Device parameters

| 设备 | 容量/kW | 能量转换效率/% | 爬坡约束/% |
|---|---|---|---|
| P2G | 500 | 60 | 20 |
| GT | 1000 | 22(气转电)、72(气转热) | 20 |
| WHB | 600 | 80 | 20 |
| GB | 800 | 82 | 20 |

表 a2　储能参数
Table a2　Energy storage parameters

| 设备 | 容量/kW | 容量上、下限约束/% | 爬坡约束/% |
|---|---|---|---|
| 电储 | 450 | 10、90 | 20 |
| 热储 | 500 | 10、90 | 20 |
| 气储 | 300 | 10、90 | 20 |